# EFFECT OF AN ANTIMICROBIAL ASSOCIATION (*RESPIREND-M®*) ON PRODUCTIVE PERFORMANCE OF BROILERS*

# EFECTO DE UNA ASOCIACIÓN ANTIMICROBIANA (*RESPIREND-M®*) SOBRE EL COMPORTAMIENTO PRODUCTIVO DE POLLOS DE CARNE*

**Mario Hernán Chuy Ko, Marcial Cumpa Gavidia y Diego Andrés Martínez Patiño-Patroni****

Facultad de Zootecnia, Universidad Nacional Agraria La Molina, Lima, Perú.

**RESUMEN**: El presente estudio tuvo como objetivo evaluar el efecto de la inclusión de una asociación antimicrobiana especial (*Respirend-M®*) en la dieta de pollos machos de carne de 0 a 14 días de edad sobre la productividad y la prevención de cuadros clínicos infeccioso bacterianos. La evaluación se realizó desde el primero hasta los catorce días de edad. Se evaluaron dos tratamientos: un tratamiento control sin *Respirend-M®* y un tratamiento con antimicrobiano (300 g de *Respirend-M®* por tonelada de alimento, de 0 a 14 días de edad). Se utilizaron 398 pollos BB machos de la línea Cobb 500, distribuidos al azar en dos tratamientos y cuatro repeticiones por tratamiento. Los parámetros evaluados fueron: peso a la 2° semana, ganancia de peso, consumo de alimento, índice de conversión alimenticia y estado sanitario. Los resultados del estudio indican que la medicación vía alimento con 300 g de *Respirend-M®* por tonelada de alimento mejora significativamente ($P<0.05$) el peso vivo final, la ganancia de peso y el consumo de alimento, sin afectar el índice de conversión alimenticia en relación al grupo control.

## INTRODUCCIÓN

Una de las etapas críticas en la producción de pollos de carne es la fase inicial, desde el primer día hasta el decimocuarto día de edad, donde en forma natural existe una alta probabilidad que los pollos se infecten con enfermedades de origen bacteriano y se presenten cuadros subclínicos, (recordemos que los pollos pueden ingerir parte del alimento que cae al suelo y su organismo se colonizará con la flora bacteriana predominante en el medio ambiente). Se estima que una falla al inicio de la campaña puede causar una mortalidad hasta del 25%.

Los cuadros subclínicos respiratorios y entéricos afectan la productividad en forma significativa, incrementando la mortalidad e impidiendo que las aves expresen su máximo potencial genético. Los diferentes niveles de estrés durante esta etapa (calor, frío, alimento, vacunaciones, agua y demás) hacen que el pollo sea más susceptible a sufrir estos desafíos. Con la finalidad de controlar estos niveles de reto durante esta etapa y disminuir sus efectos negativos sobre la productividad y salud de las aves, es común incluir antibióticos en el alimento y así coadyuvar a las aves a mantener su capacidad productiva.

El objetivo de este trabajo de investigación fue evaluar el efecto profiláctico de la inclusión de una asociación antimicrobiana (*Respirend-M®*) en la dieta de pollos comerciales de carne durante las dos primeras semanas de vida y su efecto sobre la productividad.

## MATERIALES Y MÉTODOS

### Producto evaluado

Se evaluó el producto antimicrobiano *Respirend-M®* (Marca Registrada de *CKM S.A.C.*), que es una asociación de Amoxicilina y Norfloxacina, especialmente formulado para ser administrado vía alimento, como un programa profiláctico y desarrollar un elevado efecto sinérgico.

### Lugar de experimentación

La prueba experimental se llevó a cabo en la *Unidad Experimental de Avicultura* de la *Facultad de Zootecnia* de la *Universidad Nacional Agraria La Molina* durante el mes de Marzo del 2001.

** ckm@ckmperu.com

**Duración de la prueba**

La prueba fue llevada a cabo desde el día 1 al día 14 de edad.

**Dosificación**

Se evaluó una dosis de 300 g de *Respirend-M*® por tonelada de alimento.

**Animales experimentales**

Pollos BB machos comerciales de carne de la línea Cobb.

**Conformación de grupos**

Se conformaron dos grupos de evaluación: el tratamiento 1 (grupo testigo) y el tratamiento 2 (medicado con *Respirend-M*® a 300 g/TM de alimento).

Se utilizaron 4 módulos con 50 aves cada uno, las que fueron distribuidas al azar para conformar dos tratamientos y cuatro repeticiones.

**Programa de alimentación**

El programa de alimentación fue similar para ambos grupos, diferenciándose sólo en la adición del producto en evaluación. El alimento utilizado para la prueba fue una fórmula comercial.

**Parámetros evaluados**

Los parámetros evaluados fueron (1) peso vivo a los 14 días de edad, (2) ganancia de peso de 1 a 14 días de edad, (3) consumo de alimento de 1 a 14 días de edad, (4) conversión alimenticia entre el 1º y 14º día de edad, (5) mortalidad acumulada de 1 a 14 días de edad, y reacciones post-vacunales (ruidos por minuto).

**Diseño estadístico**

El diseño utilizado para el experimento fue el Diseño Completamente al Azar con dos tratamientos y cuatro repeticiones. El análisis de varianza de los datos se llevó a cabo usando el programa *Statistical Analysis System (SAS)* y la diferencia de medias se realizó usando la Prueba de Duncan.

## RESULTADOS Y DISCUSIONES

Los resultados sobre el comportamiento productivo observados en los pollos de carne alimentados con la dieta control y con la dieta suplementada con *Respirend-M*®, se presentan en el Cuadro 1.

Debido a que las aves experimentales provinieron de un mismo lote de reproductoras, el peso vivo inicial para ambos tratamientos fue el mismo.

Al final del periodo experimental, el peso vivo promedio de los pollos que recibieron el alimento suplementado con *Respirend-M*® fue significativamente mayor que el del grupo control ($P<0.05$). La diferencia observada en peso vivo al día 14 de edad fue de 46.72 g equivalente a un incremento del 11%.

El consumo de alimento fue significativamente mayor (7.8%) en las aves que recibieron el alimento con *Respirend-M*® ($P<0.05$).

La conversión alimenticia en las aves que recibieron el alimento suplementado con *Respirend-M*® mostró una mejora del 2.36%. Considerando la relevancia del alimento en la estructura del costo de producción del pollo de carne, esta reducción en la conversión alimenticia representa un beneficio significativo para el empresario avícola.

La representación gráfica de los parámetros productivos observados durante el estudio se muestra en la Figura 1.

En la Figura 2 se muestra el comportamiento del peso vivo durante las 2 primeras semanas de vida. Es posible apreciar que las mejoras ($P<0.05$) observadas en el peso de las aves que recibieron *Respirend-M*® muestran un comportamiento sostenido en el tiempo: a los 7 días de edad la mejora observada en las aves que recibieron *Respirend-M*® fue del 7.7% y a los 14 días fue del 11%.

Considerando que, en condiciones normales, un comportamiento productivo favorable en las primeras semanas de vida brinda las condiciones necesarias para obtener buenas campañas, los resultados observados resultan de gran interés desde un punto de vista práctico.

El Cuadro 2 muestra el comportamiento de las aves durante la respuesta post-vacunal a la inmunización vía ocular contra la enfermedad de Newcastle. Se presenta también la incidencia de cojeras y la mortalidad observadas durante la prueba.

Se observó una variación en la reacción post-vacunal utilizando el método de ruidos por minuto (RPM), justo en el pico de la reacción.

La presencia de animales cojos no mostró diferencias entre tratamientos. Así también. La mortalidad a los 14 días fue nula para ambos tratamientos.

La inspección post-mortem realizada a una muestra representativa de ambos tratamientos evidenció una mayor incidencia de órganos congestionados en el grupo control (sin *Respirend-M*®). Este hallazgo no tiene mayor significancia ya que el estado general de las aves fue bueno.

## CONCLUSIONES

En base a los resultados del presente estudio, se concluye que la asociación antimicrobiana evaluada (*Respirend-M*®), adicionada en el alimento de pollos de carne de 0 a 14 días de edad a razón de 300 g por tonelada de alimento, genera mejoras del orden del 12% en velocidad de crecimiento (46 g adicionales por ave) y del 2.3% en la conversión alimenticia.

Los resultados obtenidos en el presente estudio permiten concluir que el uso de la asociación antimicrobiana evaluada (*Respirend-M*®; 300 g/tonelada) se traduce en beneficios concretos en la crianza comercial de pollos de carne.

## REFERENCIAS

**Cuadro 1. Comportamiento productivo de pollos de carne alimentados con una dieta estándar (T1) y con una dieta suplementada con *Respirend-M*® de 0 a 14 días de edad (T2).**

| Parámetro | Tratamiento 1 (Control) | | Tratamiento 2 (*Respirend-M*®) | |
|---|---|---|---|---|
| Peso vivo al día 1 de edad, g | 44 | a | 44 | a |
| Peso vivo al día 7 de edad, g | 167 | b | 180 | a |
| Peso vivo al día 14 de edad, g | 420,41 | b | 467,13 | a |
| Ganancia de peso al día 14 de edad, g | 376,41 | b | 423,12 | a |
| Consumo de alimento al día 14 de edad, g | 534,38 | b | 576,54 | a |
| Conversión alimenticia al día 14 de edad, g | 1,27 | | 1,24 | |

*a, b Promedios en una misma línea con diferentes letras son diferentes significativamente (P<0.05).*

**Figura 1. Comportamiento productivo de pollos de carne alimentados con una dieta estándar (T1) y con una dieta suplementada con *Respirend-M*® de 0 a 14 días de edad (T2).**

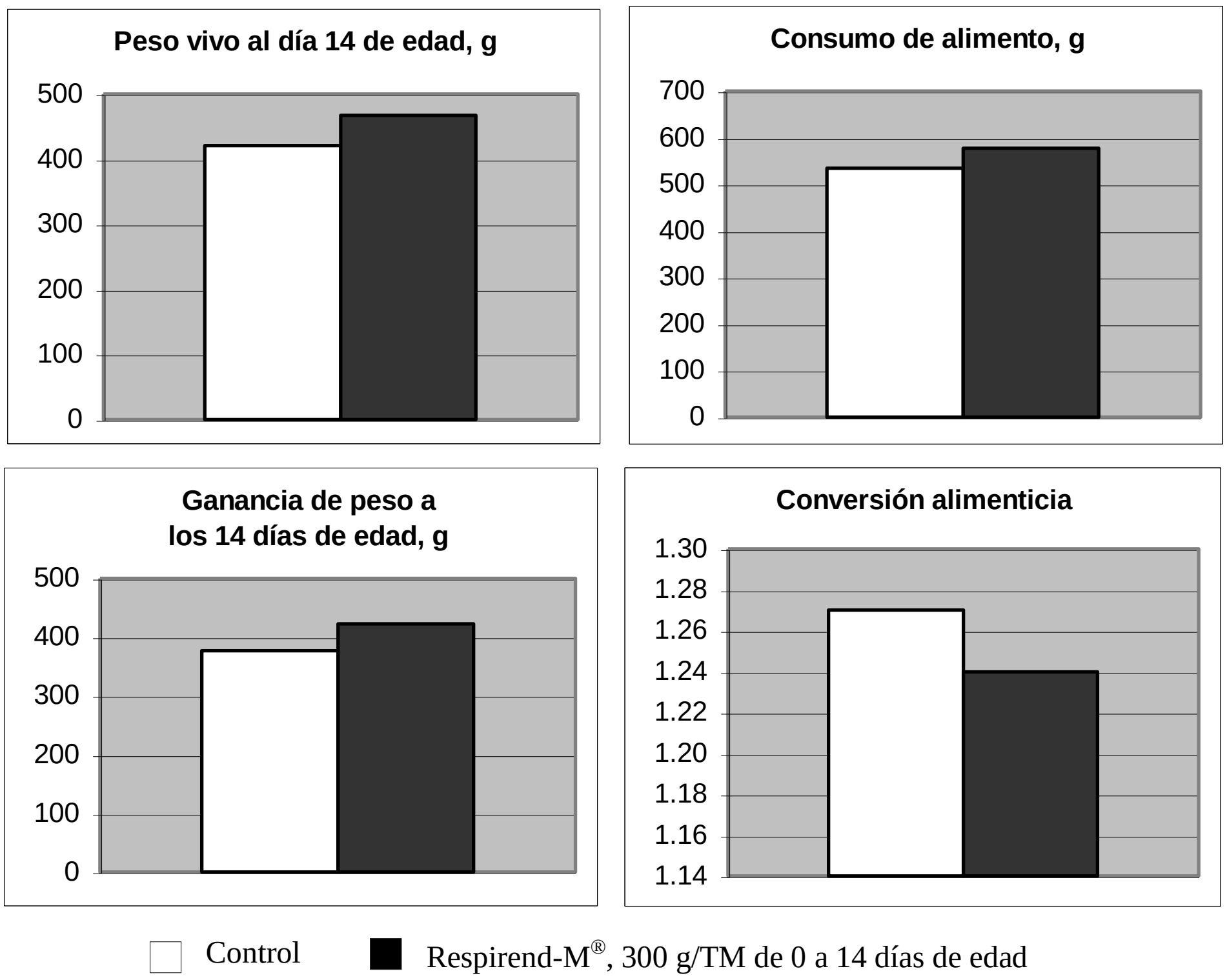

**Figura 2. Peso vivo de pollos de carne alimentados con una dieta estándar (Control) o con una dieta suplementada con *Respirend-M*® de 0 a 14 días de edad.**

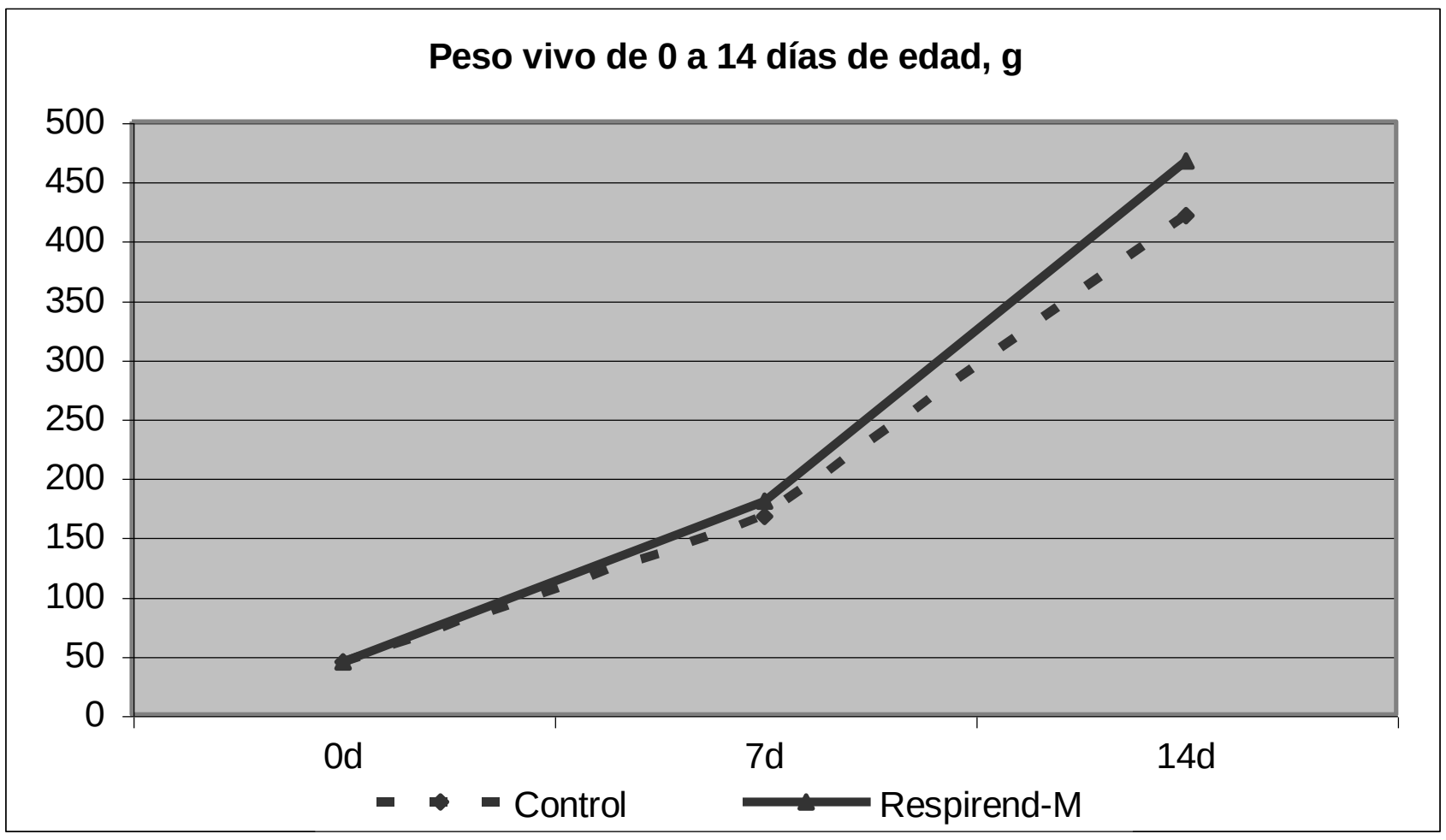

**Cuadro 2. Observaciones realizadas durante la prueba**

| Parámetro | Tratamiento 1 (Control) | Tratamiento 2 (*Respirend-M*®) |
|---|---|---|
| Reacción post-vacunal, RPM* | 17 | 7 |
| Presencia de cojos, % | 4 | 4 |
| Mortalidad, % | 0 | 0 |

* RPM: ruidos por minuto